\documentstyle[aps,multicol]{revtex}
\begin{document}
\draft
\title{ Microscopic structure of a vortex line in 
superfluid neutron star matter 
}
\author{F.V. De Blasio and \O. Elgar\o y}  
\address{ Department of Physics, University of Oslo, N-0316 Oslo, Norway \\
}
\date{\today}

\maketitle

\begin{abstract}
The microscopic structure of an isolated vortex line in superfluid 
neutron matter is studied by solving the Bogoliubov-de Gennes 
equations.  Our calculation, which is the starting point for a 
microscopic calculation of pinning forces in neutron stars, 
shows that the size of the vortex core varies differently with 
density, and is in general smaller than assumed in 
some earlier calculations of vortex pinning 
in neutron star crusts.   The implications of this result are 
discussed. 
\end{abstract}
\pacs{PACS numbers: 26.60.+c, 21.65.+f, 97.60.Jd, 97.60.Gb, 74.26.Bt}

\begin{multicols}{2}

The interior of a neutron star constitutes the only known physical system 
close to infinite nuclear and neutron matter. 
Most efforts to describe this system have mainly focussed on the 
role of nucleonic interactions at zero and finite 
temperature, while less attention has been paid to a 
microscopic description of excited states of the system,  
including superfluid vortex lines induced by the 
rotational state of the star. On the other hand, there are 
important observables of astrophysical relevance that might 
be influenced by the presence of vortex lines. 
A mutual check of the nuclear many body physics 
and of the theory of neutron star interiors comes for example  
from the study of pulsar glitches, sudden increases in the 
spinning frequency of the crust of the pulsar, followed by 
a slower tendency to conditions close to the original ones.   
It is thought that glitching events represent a  
direct manifestation of the presence of superfluid vortices  
in the interior of the star  \cite{anderson},    
the triggering event being  
an unbalance between the hydrodynamical forces acting on the 
 vortex  and the force of interaction of the 
vortex with the nuclei present in the crust (pinning force).  
There have been quite large uncertainties   
regarding the value of the pinning force, leaving room for 
quite opposite views about the validity of the 
vortex pinning model \cite{jones}. 
One source of uncertainty is related to the 
value of the pairing energy gap in uniform neutron matter, a quantity 
strongly dependent on the value of the neutron particle-particle matrix 
element near the Fermi surface \cite{wambach93,elg96,schulze96}. 
A second problem  
 is due to the very approximate way of treating vortex states 
in neutron matter. Usually a vortex is seen as a cylinder  
of normal matter (the vortex core) of radius equal 
to the BCS coherence length 
$\xi_0\approx 0.8\hbar^2 k_F / 2 m \Delta $ where 
$k_F$ is the Fermi wavenumber, $m$ is the neutron mass and $\Delta$ 
is the neutron pairing gap \cite{alpar77,taka84,gianluca95}.  
The pinning of the 
vortex to the nucleus, also treated as a  
 classical object,  is due to the 
loss of superfluidity that occurs when the two objects superimpose. 

Several experiments and theoretical 
calculations in type II superconductors  show that 
a vortex is a quite complicated quantum object \cite{general}. 
Bound states that can be formed in the center of the core as revealed by 
the increase in the     
density of states (investigated by scanning-tunneling 
microscopy) may change 
the local pairing gap in the core and the vortex size. In 
addition, they show that superfluidity is not completely 
suppressed in the core. The latter result can be in principle predicted 
also with the Ginzburg-Landau theory (GL), but unfortunately 
the GL theory is not 
reliable far from the transition temperature, which is the case 
occurring in neutron stars.   
 
In the present Letter we propose to study the structure of 
a vortex in superfluid neutron matter using a microscopic,  fully 
quantum-mechanical approach. More specifically, we shall make use of  
the Bogoliubov-de Gennes  
equations \cite{degennes}, that have been successfully employed to 
study  vortices in type II superconductors \cite{gygi,jap} and 
more in general non-homogeneous superconductivity. The use of a model 
well-tested in the laboratory represents, in our view, a major 
merit of this kind of approach. There are however important 
differences between vortices in neutron matter and those 
in superconductors as 
studied in Ref. \cite{gygi,jap} that make a direct application 
of the previous findings impossible, and require a new {\it ab-initio} 
calculation. The first obvious difference is in the basic constants 
setting the scale of the problem, like the 
mass of the particles  and the 
  density regime. The difference is not merely quantitative, since 
in our case the parameter $\xi_0 k_F$ defining the importance 
of quantum effects is smaller than in the type-II  superconductors, 
resulting in the possibility of strong deviations from previous 
solid-state calculations and also from semiclassical methods.    
Secondly, due to the charge neutrality of neutrons, magnetic fields 
do not play any role in our calculation and vortices are  
generated by the rotational  state of the star.  
A further difference is that    
the temperature in superconductors is an important variable of 
the problem, while in our case, due to the high value of the 
Fermi temperature compared to the interior temperature of a 
neutron star, it can be set equal to zero. A fourth major difference 
is the role of the inter-particle interaction, which  
in the case of neutron matter is strongly 
density-dependent. Finally, we shall be mostly interested in 
information about the pinning force, which depends in a delicate manner 
on the length scale of the vortex core.  

A vortex  state  in neutron matter can  be described 
by the Bogoliubov-de Gennes (BdG) equations \cite{degennes}: 
\begin{eqnarray} 
{\biggl ( }{{\bf p}^2 \over 2 m } +W({\bf r}) - E_F {\biggr )} 
U_{i}({\bf r}) +\Delta({\bf r}) V_i({\bf r}) &=& E_i   U_i({\bf r}) 
\label{eq:bog1} \\ 
-{\biggl ( }{{\bf p}^2\over 2 m } +W({\bf r}) - E_F {\biggr )} 
U_{i}({\bf r}) +\Delta^*({\bf r}) V_i({\bf r}) &=& E_i V_i({\bf r}) 
\label{eq:bog2}  
\end{eqnarray}
where ${\bf p}=-i\hbar\nabla$, 
$U$ and $V$ are the quasiparticle amplitudes, 
$E_F$ is the Fermi energy, 
$\Delta({\bf r})$ is the (space-dependent) pairing gap, $m$ is 
the neutron mass, $W({\bf r})$ is an external potential 
due to e.g. a nucleus, and the 
subscript $i$ represents all relevant quantum numbers.     
We write the quasiparticle states as 
\begin{eqnarray} 
U_i({\bf r})&=&{1 \over L^{1/2} }u_{n \mu}(\rho) 
\exp[i(\mu-1/2)\theta] \exp[i(k_z z)]  \label{eq:uampl} \\
V_i({\bf r})&=&{1 \over L^{1/2} }u_{n \mu}(\rho) 
\exp[i(\mu+1/2)\theta] \exp[i(k_z z)]. \label{eq:vampl}  
\end{eqnarray}
Here $L$ is the length of the cylinder, taken equal to 1 in our calculations 
since all quantities are per unit length of the vortex, $n$ is a 
radial quantum number and $\mu$ is half an odd integer.  The quantity 
$k_z$ is the (conserved) momentum in the $z$ direction.  
The pairing gap has to be calculated self-consistently as 
\begin{equation} 
\Delta({\bf r})=-g\sum_{i;0<E_i<\hbar\Omega} U_i({\bf r}) V^*_i({\bf r})
\label{eq:pairpot} 
\end{equation} 
where the sum is over quasiparticle states 
having energy $E_i$ smaller than a cutoff $\hbar\Omega$, and $g>0$ is the 
pairing strength. The meaning of these parameters is quite 
different for neutron matter compared with the solid state case, and will 
be discussed later on. Due to the above angular dependence 
of the quasiparticle states, the pairing gap  
has the form $\Delta({\bf r})=\Delta(\rho)\exp[-i\theta]$, corresponding 
to a vortex with one quantum of circulation.     
Following \cite{gygi} we expand the quasiparticle states in terms of 
cylindrically symmetric Bessel functions and impose the 
boundary condition of zero gap at the edges 
of a cylinder of radius $R$ that is chosen sufficiently large.  
More specifically, the basis functions are chosen as 
\begin{equation}
\phi_{jm}(\rho)=\frac{\sqrt{2}}{RJ_{m+1}(\alpha_{jm})}J_m\left(
\alpha_{jm}\frac{\rho}{R}\right), 
\;\; j=1,\ldots,N,
\label{eq:basis}
\end{equation}
where $m=\mu\pm \frac{1}{2}$ is an integer and $\alpha_{jm}$ is 
the $j$th zero of the Bessel function $J_m(x)$.  The dimension $N$ 
of the basis is chosen large enough to ensure convergence and 
stability of the quantities of interest.   
The quasiparticle amplitudes are expanded as 
\begin{eqnarray}
u_n(\rho)&=&\sum_j c_{nj}\phi_{j\mu-\frac{1}{2}}(\rho) \label{eq:uexp} \\
v_n(\rho)&=&\sum_j d_{nj}\phi_{j\mu+\frac{1}{2}}(\rho). \label{eq:vexp} 
\end{eqnarray}
For a given value of $\mu$ Eqs. (\ref{eq:bog1},\ref{eq:bog2}) can 
then be written as a $2N\times 2N$ matrix eigenvalue problem 
\begin{equation}
\left( \begin{array}{cc} 
       T^- & \Delta \\
       \Delta^T & T^+ 
 \end{array}\right) \Psi_n=E_n\Psi_n, 
\label{eq:eigval}
\end{equation}
where the superscript $T$ denotes the transpose of a matrix, 
$\Psi_n^T=(c_{n1},\ldots,c_{nN},d_{n1},\ldots,d_{nN})$, 
\[
T^{\pm}=\frac{\hbar^2}{2m^*}\left(\frac{\alpha_{j\mu\pm 1/2}^2}
{R^2}+k_z^2-k_F^2\right)\delta_{jj'},
\]
and the matrix $\Delta$ is given by 
\[
\Delta_{jj'}=\int_0^R \phi_{j\mu-1/2}(\rho)\Delta(\rho)\phi_{j'\mu+1/2}
(\rho)\rho d\rho.
\]   
We have in these equations set the mean field $W({\bf r})$ 
equal to zero, corresponding to an isolated vortex line in 
infinite neutron matter.   Interactions between 
neutrons set up an internal mean field which can be simulated 
by replacing the neutron mass $m$ by an effective mass $m^*$.  
However, for dilute neutron matter $m^*\approx m$ 
\cite{elg96}.
Starting from an appropriate approximation to $\Delta(\rho)$, we 
solve Eq. (\ref{eq:eigval}) for several values of $\mu$ to obtain 
the quasiparticle amplitudes $U$ and $V$.  A new approximation 
to $\Delta(\rho)$ is then obtained from Eq. (\ref{eq:pairpot}).  
The procedure is iterated until convergence is achieved.  

The BdG equations as written above have been specialized to 
the case of a point-like neutron-neutron interaction of the form 
$v({\bf r},{\bf r}')= - g \delta({\bf r} - {\bf r}') $. This 
is the same approximation as used in Refs. \cite{degennes,gygi,jap} 
and should be considered even more reliable in the case of neutron 
matter, where the range of the inter-particle interaction is of 
the order $\sim 1\;{\rm fm}$, and is thus larger than the length 
scale of the inhomogeneities, which is of the order of several fm. In 
dealing with a zero-range force, a cutoff energy $\hbar\Omega$ 
has to be introduced for the gap equation to converge. The 
density dependence of the pairing gap as calculated 
with finite-range forces, see for example  \cite{elg96}, 
can be easily translated to a dependence of $ g $ on the 
density. We shall thus fix  $\hbar \Omega = 50\;{\rm MeV}$ and 
use the value of $ g $ that reproduces the gap at large distance
from the vortex core as calculated  for a homogeneous system.    

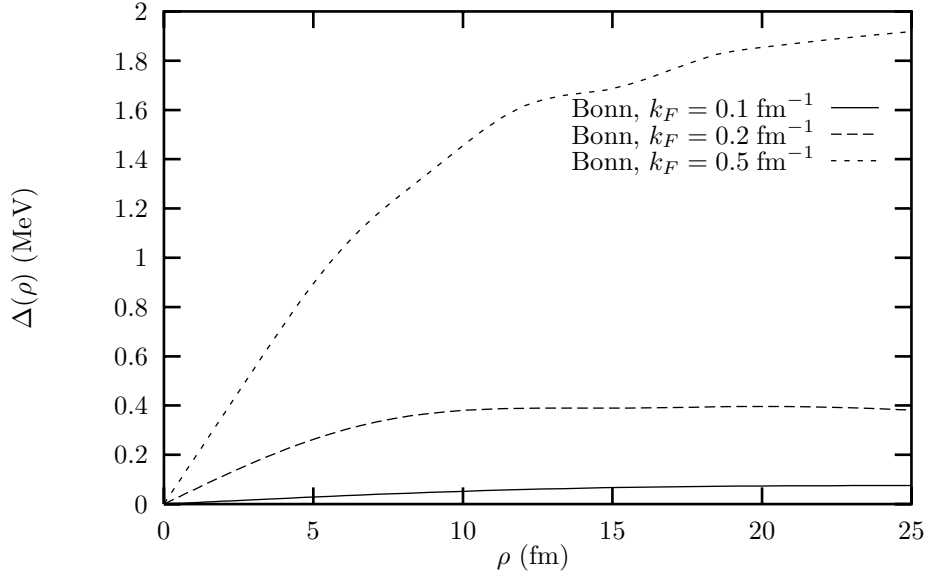
\begin{figure}
\setlength{\unitlength}{0.1bp}
\special{!
/gnudict 40 dict def
gnudict begin
/Color false def
/Solid false def
/gnulinewidth 5.000 def
/vshift -33 def
/dl {10 mul} def
/hpt 31.5 def
/vpt 31.5 def
/M {moveto} bind def
/L {lineto} bind def
/R {rmoveto} bind def
/V {rlineto} bind def
/vpt2 vpt 2 mul def
/hpt2 hpt 2 mul def
/Lshow { currentpoint stroke M
  0 vshift R show } def
/Rshow { currentpoint stroke M
  dup stringwidth pop neg vshift R show } def
/Cshow { currentpoint stroke M
  dup stringwidth pop -2 div vshift R show } def
/DL { Color {setrgbcolor Solid {pop []} if 0 setdash }
 {pop pop pop Solid {pop []} if 0 setdash} ifelse } def
/BL { stroke gnulinewidth 2 mul setlinewidth } def
/AL { stroke gnulinewidth 2 div setlinewidth } def
/PL { stroke gnulinewidth setlinewidth } def
/LTb { BL [] 0 0 0 DL } def
/LTa { AL [1 dl 2 dl] 0 setdash 0 0 0 setrgbcolor } def
/LT0 { PL [] 0 1 0 DL } def
/LT1 { PL [4 dl 2 dl] 0 0 1 DL } def
/LT2 { PL [2 dl 3 dl] 1 0 0 DL } def
/LT3 { PL [1 dl 1.5 dl] 1 0 1 DL } def
/LT4 { PL [5 dl 2 dl 1 dl 2 dl] 0 1 1 DL } def
/LT5 { PL [4 dl 3 dl 1 dl 3 dl] 1 1 0 DL } def
/LT6 { PL [2 dl 2 dl 2 dl 4 dl] 0 0 0 DL } def
/LT7 { PL [2 dl 2 dl 2 dl 2 dl 2 dl 4 dl] 1 0.3 0 DL } def
/LT8 { PL [2 dl 2 dl 2 dl 2 dl 2 dl 2 dl 2 dl 4 dl] 0.5 0.5 0.5 DL } def
/P { stroke [] 0 setdash
  currentlinewidth 2 div sub M
  0 currentlinewidth V stroke } def
/D { stroke [] 0 setdash 2 copy vpt add M
  hpt neg vpt neg V hpt vpt neg V
  hpt vpt V hpt neg vpt V closepath stroke
  P } def
/A { stroke [] 0 setdash vpt sub M 0 vpt2 V
  currentpoint stroke M
  hpt neg vpt neg R hpt2 0 V stroke
  } def
/B { stroke [] 0 setdash 2 copy exch hpt sub exch vpt add M
  0 vpt2 neg V hpt2 0 V 0 vpt2 V
  hpt2 neg 0 V closepath stroke
  P } def
/C { stroke [] 0 setdash exch hpt sub exch vpt add M
  hpt2 vpt2 neg V currentpoint stroke M
  hpt2 neg 0 R hpt2 vpt2 V stroke } def
/T { stroke [] 0 setdash 2 copy vpt 1.12 mul add M
  hpt neg vpt -1.62 mul V
  hpt 2 mul 0 V
  hpt neg vpt 1.62 mul V closepath stroke
  P  } def
/S { 2 copy A C} def
end
}
\begin{picture}(3600,2160)(0,0)
\special{"
gnudict begin
gsave
50 50 translate
0.100 0.100 scale
0 setgray
/Helvetica findfont 100 scalefont setfont
newpath
-500.000000 -500.000000 translate
LTa
600 251 M
2817 0 V
600 251 M
0 1858 V
LTb
600 251 M
63 0 V
2754 0 R
-63 0 V
600 437 M
63 0 V
2754 0 R
-63 0 V
600 623 M
63 0 V
2754 0 R
-63 0 V
600 808 M
63 0 V
2754 0 R
-63 0 V
600 994 M
63 0 V
2754 0 R
-63 0 V
600 1180 M
63 0 V
2754 0 R
-63 0 V
600 1366 M
63 0 V
2754 0 R
-63 0 V
600 1552 M
63 0 V
2754 0 R
-63 0 V
600 1737 M
63 0 V
2754 0 R
-63 0 V
600 1923 M
63 0 V
2754 0 R
-63 0 V
600 2109 M
63 0 V
2754 0 R
-63 0 V
600 251 M
0 63 V
0 1795 R
0 -63 V
1163 251 M
0 63 V
0 1795 R
0 -63 V
1727 251 M
0 63 V
0 1795 R
0 -63 V
2290 251 M
0 63 V
0 1795 R
0 -63 V
2854 251 M
0 63 V
0 1795 R
0 -63 V
3417 251 M
0 63 V
0 1795 R
0 -63 V
600 251 M
2817 0 V
0 1858 V
-2817 0 V
600 251 L
LT0
3114 1746 M
180 0 V
601 251 M
3 0 V
6 1 V
9 0 V
12 1 V
14 0 V
18 1 V
19 1 V
23 1 V
25 1 V
28 2 V
30 1 V
33 2 V
35 1 V
38 2 V
41 2 V
42 2 V
46 2 V
47 2 V
50 3 V
52 2 V
54 2 V
57 3 V
58 2 V
61 3 V
62 2 V
65 2 V
66 3 V
68 2 V
70 2 V
71 3 V
73 2 V
75 2 V
76 2 V
77 1 V
78 2 V
80 2 V
81 1 V
82 1 V
83 1 V
84 1 V
84 1 V
86 1 V
86 0 V
86 1 V
87 0 V
87 0 V
88 1 V
88 0 V
88 0 V
44 0 V
LT1
3114 1646 M
180 0 V
601 251 M
3 2 V
6 3 V
9 4 V
12 6 V
14 7 V
18 8 V
19 10 V
23 10 V
25 12 V
28 14 V
30 14 V
33 15 V
35 16 V
38 17 V
41 18 V
42 18 V
46 18 V
47 19 V
50 18 V
52 18 V
54 17 V
57 17 V
58 15 V
61 14 V
62 12 V
65 11 V
66 8 V
68 7 V
70 6 V
71 3 V
73 3 V
75 1 V
76 1 V
77 0 V
78 0 V
80 0 V
81 0 V
82 1 V
83 1 V
84 1 V
84 2 V
86 1 V
86 0 V
86 0 V
87 -1 V
87 -2 V
88 -2 V
88 -3 V
88 -4 V
44 -1 V
LT2
3114 1546 M
180 0 V
601 252 M
3 5 V
6 10 V
9 13 V
12 18 V
14 21 V
18 26 V
19 30 V
23 34 V
25 37 V
28 42 V
30 46 V
33 49 V
35 53 V
38 57 V
41 61 V
42 64 V
46 67 V
47 69 V
50 71 V
52 69 V
54 68 V
57 65 V
58 60 V
61 56 V
62 52 V
65 51 V
66 52 V
68 54 V
70 56 V
71 54 V
73 49 V
75 40 V
76 25 V
77 16 V
78 10 V
80 10 V
81 19 V
82 26 V
83 31 V
84 30 V
84 28 V
86 16 V
86 11 V
86 10 V
87 10 V
87 10 V
88 9 V
88 9 V
88 8 V
44 4 V
stroke
grestore
end
showpage
}
\put(3054,1546){\makebox(0,0)[r]{Bonn, $k_F=0.5\;{\rm fm}^{-1}$}}
\put(3054,1646){\makebox(0,0)[r]{Bonn, $k_F=0.2\;{\rm fm}^{-1}$}}
\put(3054,1746){\makebox(0,0)[r]{Bonn, $k_F=0.1\;{\rm fm}^{-1}$}}
\put(2008,51){\makebox(0,0){$\rho$ (${\rm fm}$) }}
\put(100,1180){%
\special{ps: gsave currentpoint currentpoint translate
270 rotate neg exch neg exch translate}%
\makebox(0,0)[b]{\shortstack{$\Delta(\rho)$ (${\rm MeV}$) }}%
\special{ps: currentpoint grestore moveto}%
}
\put(3417,151){\makebox(0,0){25}}
\put(2854,151){\makebox(0,0){20}}
\put(2290,151){\makebox(0,0){15}}
\put(1727,151){\makebox(0,0){10}}
\put(1163,151){\makebox(0,0){5}}
\put(600,151){\makebox(0,0){0}}
\put(540,2109){\makebox(0,0)[r]{2}}
\put(540,1923){\makebox(0,0)[r]{1.8}}
\put(540,1737){\makebox(0,0)[r]{1.6}}
\put(540,1552){\makebox(0,0)[r]{1.4}}
\put(540,1366){\makebox(0,0)[r]{1.2}}
\put(540,1180){\makebox(0,0)[r]{1}}
\put(540,994){\makebox(0,0)[r]{0.8}}
\put(540,808){\makebox(0,0)[r]{0.6}}
\put(540,623){\makebox(0,0)[r]{0.4}}
\put(540,437){\makebox(0,0)[r]{0.2}}
\put(540,251){\makebox(0,0)[r]{0}}
\end{picture}
	\caption{Pairing potential $\Delta(\rho)$ as a function of 
$\rho$ for an isolated vortex in neutron matter.}
    \label{fig:fig1}
\end{figure}

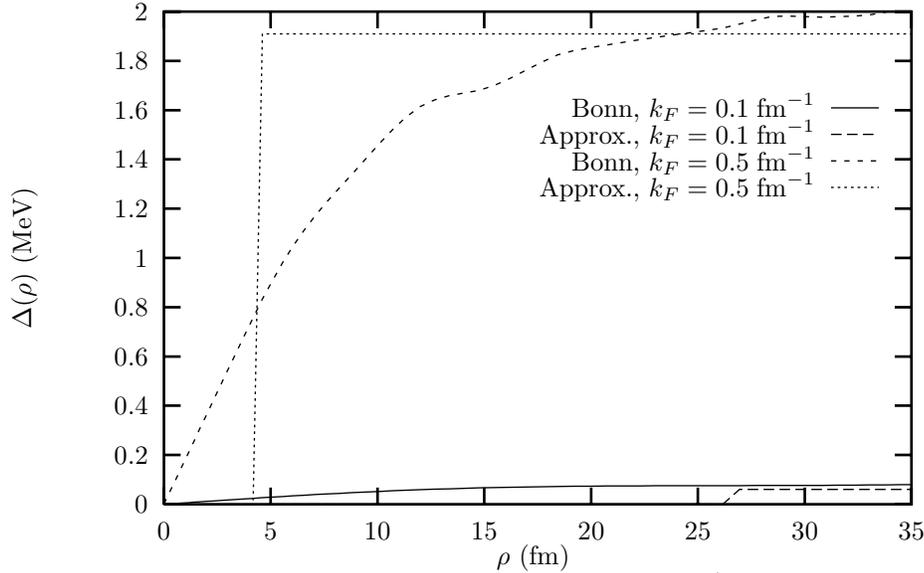
\begin{figure}
\setlength{\unitlength}{0.1bp}
\special{!
/gnudict 40 dict def
gnudict begin
/Color false def
/Solid false def
/gnulinewidth 5.000 def
/vshift -33 def
/dl {10 mul} def
/hpt 31.5 def
/vpt 31.5 def
/M {moveto} bind def
/L {lineto} bind def
/R {rmoveto} bind def
/V {rlineto} bind def
/vpt2 vpt 2 mul def
/hpt2 hpt 2 mul def
/Lshow { currentpoint stroke M
  0 vshift R show } def
/Rshow { currentpoint stroke M
  dup stringwidth pop neg vshift R show } def
/Cshow { currentpoint stroke M
  dup stringwidth pop -2 div vshift R show } def
/DL { Color {setrgbcolor Solid {pop []} if 0 setdash }
 {pop pop pop Solid {pop []} if 0 setdash} ifelse } def
/BL { stroke gnulinewidth 2 mul setlinewidth } def
/AL { stroke gnulinewidth 2 div setlinewidth } def
/PL { stroke gnulinewidth setlinewidth } def
/LTb { BL [] 0 0 0 DL } def
/LTa { AL [1 dl 2 dl] 0 setdash 0 0 0 setrgbcolor } def
/LT0 { PL [] 0 1 0 DL } def
/LT1 { PL [4 dl 2 dl] 0 0 1 DL } def
/LT2 { PL [2 dl 3 dl] 1 0 0 DL } def
/LT3 { PL [1 dl 1.5 dl] 1 0 1 DL } def
/LT4 { PL [5 dl 2 dl 1 dl 2 dl] 0 1 1 DL } def
/LT5 { PL [4 dl 3 dl 1 dl 3 dl] 1 1 0 DL } def
/LT6 { PL [2 dl 2 dl 2 dl 4 dl] 0 0 0 DL } def
/LT7 { PL [2 dl 2 dl 2 dl 2 dl 2 dl 4 dl] 1 0.3 0 DL } def
/LT8 { PL [2 dl 2 dl 2 dl 2 dl 2 dl 2 dl 2 dl 4 dl] 0.5 0.5 0.5 DL } def
/P { stroke [] 0 setdash
  currentlinewidth 2 div sub M
  0 currentlinewidth V stroke } def
/D { stroke [] 0 setdash 2 copy vpt add M
  hpt neg vpt neg V hpt vpt neg V
  hpt vpt V hpt neg vpt V closepath stroke
  P } def
/A { stroke [] 0 setdash vpt sub M 0 vpt2 V
  currentpoint stroke M
  hpt neg vpt neg R hpt2 0 V stroke
  } def
/B { stroke [] 0 setdash 2 copy exch hpt sub exch vpt add M
  0 vpt2 neg V hpt2 0 V 0 vpt2 V
  hpt2 neg 0 V closepath stroke
  P } def
/C { stroke [] 0 setdash exch hpt sub exch vpt add M
  hpt2 vpt2 neg V currentpoint stroke M
  hpt2 neg 0 R hpt2 vpt2 V stroke } def
/T { stroke [] 0 setdash 2 copy vpt 1.12 mul add M
  hpt neg vpt -1.62 mul V
  hpt 2 mul 0 V
  hpt neg vpt 1.62 mul V closepath stroke
  P  } def
/S { 2 copy A C} def
end
}
\begin{picture}(3600,2160)(0,0)
\special{"
gnudict begin
gsave
50 50 translate
0.100 0.100 scale
0 setgray
/Helvetica findfont 100 scalefont setfont
newpath
-500.000000 -500.000000 translate
LTa
600 251 M
2817 0 V
600 251 M
0 1858 V
LTb
600 251 M
63 0 V
2754 0 R
-63 0 V
600 437 M
63 0 V
2754 0 R
-63 0 V
600 623 M
63 0 V
2754 0 R
-63 0 V
600 808 M
63 0 V
2754 0 R
-63 0 V
600 994 M
63 0 V
2754 0 R
-63 0 V
600 1180 M
63 0 V
2754 0 R
-63 0 V
600 1366 M
63 0 V
2754 0 R
-63 0 V
600 1552 M
63 0 V
2754 0 R
-63 0 V
600 1737 M
63 0 V
2754 0 R
-63 0 V
600 1923 M
63 0 V
2754 0 R
-63 0 V
600 2109 M
63 0 V
2754 0 R
-63 0 V
600 251 M
0 63 V
0 1795 R
0 -63 V
1002 251 M
0 63 V
0 1795 R
0 -63 V
1405 251 M
0 63 V
0 1795 R
0 -63 V
1807 251 M
0 63 V
0 1795 R
0 -63 V
2210 251 M
0 63 V
0 1795 R
0 -63 V
2612 251 M
0 63 V
0 1795 R
0 -63 V
3015 251 M
0 63 V
0 1795 R
0 -63 V
3417 251 M
0 63 V
0 1795 R
0 -63 V
600 251 M
2817 0 V
0 1858 V
-2817 0 V
600 251 L
LT0
3114 1746 M
180 0 V
601 251 M
2 0 V
4 1 V
7 0 V
8 1 V
10 0 V
13 1 V
14 1 V
16 1 V
18 1 V
20 2 V
21 1 V
24 2 V
25 1 V
27 2 V
29 2 V
31 2 V
32 2 V
34 2 V
35 3 V
38 2 V
38 2 V
41 3 V
42 2 V
43 3 V
44 2 V
46 2 V
48 3 V
48 2 V
50 2 V
51 3 V
52 2 V
54 2 V
54 2 V
55 1 V
56 2 V
57 2 V
58 1 V
58 1 V
60 1 V
59 1 V
61 1 V
61 1 V
61 0 V
62 1 V
62 0 V
62 0 V
63 1 V
63 0 V
63 0 V
63 0 V
62 0 V
63 0 V
63 1 V
62 0 V
62 0 V
62 0 V
62 0 V
61 1 V
60 0 V
60 1 V
59 0 V
59 1 V
38 0 V
LT1
3114 1646 M
180 0 V
601 251 M
2 0 V
4 0 V
7 0 V
8 0 V
10 0 V
13 0 V
14 0 V
16 0 V
18 0 V
20 0 V
21 0 V
24 0 V
25 0 V
27 0 V
29 0 V
31 0 V
32 0 V
34 0 V
35 0 V
38 0 V
38 0 V
41 0 V
42 0 V
43 0 V
44 0 V
46 0 V
48 0 V
48 0 V
50 0 V
51 0 V
52 0 V
54 0 V
54 0 V
55 0 V
56 0 V
57 0 V
58 0 V
58 0 V
60 0 V
59 0 V
61 0 V
61 0 V
61 0 V
62 0 V
62 0 V
62 0 V
63 0 V
63 0 V
63 0 V
63 0 V
62 0 V
63 56 V
63 0 V
62 0 V
62 0 V
62 0 V
62 0 V
61 0 V
60 0 V
60 0 V
59 0 V
59 0 V
38 0 V
LT2
3114 1546 M
180 0 V
601 252 M
2 5 V
4 10 V
7 13 V
8 18 V
10 21 V
13 26 V
14 30 V
16 34 V
18 37 V
20 42 V
21 46 V
24 49 V
25 53 V
27 57 V
29 61 V
31 64 V
32 67 V
34 69 V
35 71 V
38 69 V
38 68 V
41 65 V
42 60 V
43 56 V
44 52 V
46 51 V
48 52 V
48 54 V
50 56 V
51 54 V
52 49 V
54 40 V
54 25 V
55 16 V
56 10 V
57 10 V
58 19 V
58 26 V
60 31 V
59 30 V
61 28 V
61 16 V
61 11 V
62 10 V
62 10 V
62 10 V
63 9 V
63 9 V
63 8 V
63 9 V
62 9 V
63 15 V
63 16 V
62 13 V
62 1 V
62 -2 V
62 -2 V
61 2 V
60 2 V
60 5 V
59 10 V
9 2 V
LT3
3114 1446 M
180 0 V
601 251 M
2 0 V
4 0 V
7 0 V
8 0 V
10 0 V
13 0 V
14 0 V
16 0 V
18 0 V
20 0 V
21 0 V
24 0 V
25 0 V
27 0 V
29 0 V
31 0 V
32 0 V
34 0 V
35 1774 V
38 0 V
38 0 V
41 0 V
42 0 V
43 0 V
44 0 V
46 0 V
48 0 V
48 0 V
50 0 V
51 0 V
52 0 V
54 0 V
54 0 V
55 0 V
56 0 V
57 0 V
58 0 V
58 0 V
60 0 V
59 0 V
61 0 V
61 0 V
61 0 V
62 0 V
62 0 V
62 0 V
63 0 V
63 0 V
63 0 V
63 0 V
62 0 V
63 0 V
63 0 V
62 0 V
62 0 V
62 0 V
62 0 V
61 0 V
60 0 V
60 0 V
59 0 V
59 0 V
38 0 V
stroke
grestore
end
showpage
}
\put(3054,1446){\makebox(0,0)[r]{Approx., $k_F=0.5\;{\rm fm}^{-1}$}}
\put(3054,1546){\makebox(0,0)[r]{Bonn, $k_F=0.5\;{\rm fm}^{-1}$}}
\put(3054,1646){\makebox(0,0)[r]{Approx., $k_F=0.1\;{\rm fm}^{-1}$}}
\put(3054,1746){\makebox(0,0)[r]{Bonn, $k_F=0.1\;{\rm fm}^{-1}$}}
\put(2008,51){\makebox(0,0){$\rho$ (${\rm fm}$) }}
\put(100,1180){%
\special{ps: gsave currentpoint currentpoint translate
270 rotate neg exch neg exch translate}%
\makebox(0,0)[b]{\shortstack{$\Delta(\rho)$ (${\rm MeV}$) }}%
\special{ps: currentpoint grestore moveto}%
}
\put(3417,151){\makebox(0,0){35}}
\put(3015,151){\makebox(0,0){30}}
\put(2612,151){\makebox(0,0){25}}
\put(2210,151){\makebox(0,0){20}}
\put(1807,151){\makebox(0,0){15}}
\put(1405,151){\makebox(0,0){10}}
\put(1002,151){\makebox(0,0){5}}
\put(600,151){\makebox(0,0){0}}
\put(540,2109){\makebox(0,0)[r]{2}}
\put(540,1923){\makebox(0,0)[r]{1.8}}
\put(540,1737){\makebox(0,0)[r]{1.6}}
\put(540,1552){\makebox(0,0)[r]{1.4}}
\put(540,1366){\makebox(0,0)[r]{1.2}}
\put(540,1180){\makebox(0,0)[r]{1}}
\put(540,994){\makebox(0,0)[r]{0.8}}
\put(540,808){\makebox(0,0)[r]{0.6}}
\put(540,623){\makebox(0,0)[r]{0.4}}
\put(540,437){\makebox(0,0)[r]{0.2}}
\put(540,251){\makebox(0,0)[r]{0}}
\end{picture}
	\caption{The exact pairing potentials at $k_F=0.1$ and 
$0.5\;{\rm fm}^{-1}$ compared with those  
obtained by using the $\xi_0$ cutoff prescription described in the text.}
    \label{fig:fig2}
\end{figure}

Fig. \ref{fig:fig1} shows the pairing gap as 
a function of the distance from the vortex axis, 
$\Delta(\rho)$  using the above formalism and for three 
different values of the neutron Fermi wavenumber. In all cases  
the gap increases  from a  value zero to an asymptotic  
one $\Delta_{\infty}$. The latter quantity represents evidently 
the value of the gap as calculated for uniform neutron matter.  
In the figure   
 we have fixed the value of the effective pairing strength $g$ to 
reproduce the value as calculated from the bare Bonn A potential 
\cite{elg96}.   
Like for vortices in type II superconductors, the gap is found 
to increase linearly for small values of $\rho$ before 
reaching the asymptotic value. The linear rate of increase of the 
gap defines a  coherence length that for consistency with 
ref. \cite{gygi} we call $\xi_2$, given by
\begin{equation}
\lim_{\rho\rightarrow 0} \Delta(\rho)=\Delta_{\infty}\frac{\rho}{\xi_2}.
\label{eq:xi2def}
\end{equation} 
As visible from the figure this coherence length  
represents an appropriate length scale of the vortex core.
 A second  length scale, the BCS coherence length $\xi_0$,   
can be  defined as the distance from the 
vortex axis at which  the  
energy gain of the Cooper pair due to the velocity field  
becomes of the order of the energy necessary to break the pair 
(or also by the condition that the  
superfluid 
velocity $v_s=\hbar/m\rho$ coincides with the Landau velocity 
for the suppression of superfluidity, $ 2\Delta_{\infty} m/\hbar k_F$), 
and it is found $\xi_0=16.88 k_F/\Delta_{\infty}$.  
This BCS coherence length, which microscopically  
represents the size of a Cooper pair \cite{noi_cohere},  
is the one usually employed to calculate  
the interaction of a vortex with the nuclei in the crust 
of a neutron star \cite{alpar77,taka84,gianluca95,alpar_pines}. 
In particular, matter is supposed to be completely normal 
($\Delta(\rho)=0$) for $\rho<\xi_0$ and superfluid 
($\Delta(\rho)=\Delta_{\infty}$) for $\rho>\xi_0$. In a 
pictorial view, this length can be obtained intersecting the 
gap at infinity, $\Delta_{\infty}$  
with the curve $ g(\rho)=0.814\times \hbar^2k_F/2 m \rho $.  
The resulting gap profile is  
shown in Fig. \ref{fig:fig2} for the cases $k_F=0.1\;{\rm fm}^{-1}$ and 
$k_F=0.5\;{\rm fm}^{-1}$.   
It is evident that this definition can be a reliable 
estimate of the vortex core size only when the intersection 
occurs for values of $\rho$ smaller than $\xi_2$. The plot  
for $k_F=0.1\;{\rm fm}^{-1}$ is a case  where the intersection occurs 
where the gap  has already saturated. 
According to the common view of neutron star pinning, 
all the matter at $\rho<\xi_0$ should be considered in 
a normal state. This is not the case, as clearly visible 
from the figure.   For $k_F=0.5\;{\rm fm}^{-1}$, Fig. \ref{fig:fig2} 
shows that $\xi_0<\xi_2$, and the approximation above is probably 
not too bad, although it somewhat overestimates the gap in the region 
$\xi_0<\rho<\xi_2$.  

Table \ref{tab:tab1} shows $\xi_0$ and $\xi_2$ as functions of 
the Fermi momentum, where the pairing strength $g$ has been fitted 
to the $^1S_0$ gap of the Bonn A potential \cite{elg96}.  
It is visible from the table that the value of $\xi_2$ seems 
to be uncorrelated with $\xi_0$. In general, $\xi_2$ remains 
within values between 6 to 10 fm, while the values of $\xi_0$ 
are much more scattered.  
Due to the fact that Friedel oscillations 
may slightly distort the vortex structure at small 
distances from the core \cite{jap}, we  
also studied some vortex length scales 
defined without making use of the behavior close to the axis, 
for example the distance 
from the vortex core at which the gap is a fraction of 
$\Delta_{\infty}$, typically 50 \% and 90 \%.  As expected, these 
quantities are larger than $\xi_2$, but show a similar trend.       
Overall, we found these lengths to agree better with $\xi_2$ 
than with $\xi_0$.
 
To understand the scaling of the vortex size as a function of 
the Fermi momentum, we also performed calculations 
keeping the pairing strength fixed and changing only the 
Fermi wavenumber, even if this produced unrealistic gaps.  
We found $\xi_2$ to be a decreasing function of the wavenumber,   
a result which  
agrees qualitatively with previous work on vortices 
in superconductors \cite{gygi,kramer} where the vortex size is 
predicted to scale as $\sim k_F^{-1}$, but is very different 
from the behavior of the     
 BCS coherence length, which scales proportionally to $k_F/\Delta_{\infty}$.  
In the realistic case, where the pairing strength is fitted to the 
gap in uniform matter, the coherence length $\xi_2$ increases as a 
function of $k_F$.   
Thus, the density dependence of the pairing strength $g$  
is also very important, and  
the vortex size appears to depend in a quite complex way both on $g$ 
and on the Fermi wavenumber. 
Studying the distribution of the 
eigenvalues,  we have found indeed 
that several bound states are formed in the vortex core. This 
feature is different from the solid-state case \cite{gygi} 
where essentially  
only one bound state at each value of the angular momentum 
$\mu$ is formed. The value of $\xi_2$  
is probably affected in a complex way by all the bound states, 
making a simple scaling law difficult to work out. As a 
partial conclusion, however, we can state that when varying the 
Fermi wavenumber and the pairing strength over the range of 
values relevant for $^1S_0$ neutron pairing, the coherence length 
$\xi_2$ does not change by more than 50 \%, while 
$\xi_0$ can vary by a factor of ten. 
  
We are now able to speculate on the applications of our 
results to the case of vortex pinning in the crust. The 
use of $\xi_2$ instead of $\xi_0$ as the length scale of the 
vortex core, implies that the pinning regime is approximately 
the same in the whole inner crust. In particular  
the super-weak pinning, proposed in the  
case  of very large vortex  
core and in which  several nuclei may be enveloped on 
one single plane of the vortex, is at variance  
with our findings.  
A second consequence  already predictable on the 
basis of the data presented   
       is that the pinning force exerted by 
a nucleus on a vortex should scale like 
$ \sim \Delta^2 k_F/ \xi_2 $. We expect an increase of 
the pinning force of the order $\sim \xi_0/\xi_2$ compared 
with the cases where $\xi_0$ is used for the size of the vortex core, 
and this can be a very large factor for low densities.  

We also examined the distortion of a vortex line in the presence 
of a mean field potential of cylindrical symmetry. To simulate 
the presence of a nuclear cluster,  we chose a Woods-Saxon shape 
for the potential $W({\bf r})$  
with nuclear parameters of the order of those expected in the 
inner crust of a neutron star. In presence of a potential 
field, the diagonal parts of the Hamiltonian matrix become  
$T^{\pm}\rightarrow T^{\pm}+W^{\pm}$ where 
the matrix elements of the potential matrix are of the form 
$W^{\pm}_{j j'}=\int d\rho \rho \phi_{j \mu \pm 1/2} (\rho) W(\rho) 
\phi_{j' \mu \pm 1/2} (\rho) \delta_{j j'} $.  
We found a decrease 
of the coherence length in presence of the field. This shows that 
the shape of the vortex line is sensitive to the presence 
of the nucleus, an effect that might have important 
consequences in vortex pinning.  
This in fact contrasts with 
the assumption implicit in many calculations of vortex pinning 
in neutron star crusts, where the vortex core is assumed to be inert
during the pinning state.  
 
A further advantage the BdG  formalism, not exploited in the 
present study,   
 is that the Hamiltonian is  diagonalized, 
 resulting in a consistent  calculation of the 
energy, while in  other approaches 
  the local density approximation 
\cite{luciano_etal} or Ginzburg-Landau approaches \cite{baym} need 
to be used.  
The calculation of the pinning energy and force of a nucleus with 
a superfluid vortex,  
technically more involved but conceptually straightforward,   
is thus under consideration \cite{all}.

We thank N. Hayashi for providing information about the   
calculational details of Ref. \cite{jap}.  
F. V. De Blasio would also like to thank J. Paaske for fruitful 
discussions on vortices in superconductors, and NORDITA, Copenhagen 
for hospitality during the development of parts of this work.   
\O. Elgar\o y thanks A. Botnen for useful advice on  
the computational details.  
 

\end{multicols}

\newpage


\begin{table}[t]
\begin{center}
\begin{tabular}{llll} 
\multicolumn{1}{c}{$k_F\;({\rm fm}^{-1}$)}&
\multicolumn{1}{c}{$\Delta_\infty$}&
\multicolumn{1}{c}{$\xi_0$ (fm)}&\multicolumn{1}{c}{$\xi_2$ (fm)}\\ \hline 
   0.1   & 0.06  & 26.80 & 11.46      \\   
   0.2   & 0.38  & 8.73  & 6.59       \\
   0.5   & 1.91  & 4.42  & 10.20      \\ \hline 
\end{tabular}
\caption{Gap at infinity and the length scales  
$\xi_0$ and $\xi_2$ as functions of density.}
\label{tab:tab1}
\end{center}
\end{table}

\end{document}